# Superconductivity on the verge of Mott localization in ternary iron sulfide

Jiangang Guo, Xiaolong Chen[*], Gang Wang, Tingting Zhou, Xiaofang Lai, Shifeng Jin, Shunchong Wang, Kaixing Zhu

[†] Research & Development Center for Functional Crystals, Beijing National Laboratory for Condensed Matter Physics, Institute of Physics, Chinese Academy of Sciences, P.O. Box 603, Beijing 100190, China

[*]E-mail: chenx29 @aphy.iphy.ac.cn.

## Abstract

We report the results of electrical and magnetic properties on two new compounds, $K_{0.8}Fe_{1.7}S_2$ and $K_{0.8}Fe_{1.7}SeS$, both having similar structures to newly discovered superconducting $K_{0.8}Fe_{1.7}Se_2$. $K_{0.8}Fe_{1.7}S_2$ exhibits a semiconductor-like electrical property and undergoes an anti-ferromagnetic transition at about 260 K. Upon replacing half of S with Se, $K_{0.8}Fe_{1.7}SSe$ becomes a superconductor at 25 K, implying the superconductivity evolves from a Mott AFM state in Fe-Se based superconductors.

## Introduction

The discovery of superconductivity in the anti-PbO-type FeSe [1] has aroused a lot of interests in searching new iron-based superconductors due to its simplest structure and tunable superconductivity. The angle resolved photoemission spectroscopy (ARPES) showed that the normal state of $FeSe_{0.42}Te_{0.58}$ is a strongly correlated metal, which is significantly different from the 1111 and 122 iron pnictide families [2]. The partial sulfur and tellurium substitutions for Se enhanced the superconducting transition temperature up to 15.5 K and 15.2 K, respectively [3, 4]. Simultaneously, the superconductivity transition temperature of FeSe could be enhanced up to 37 K by the exertion of high pressures [5]. More recently, our work about $K_xFe_{1.7}Se_2$ demonstrated that K cations intercalation could strikingly enhance the $T_c$ up to ~30 K under ambient pressure [6]. After that, the following works confirmed that other monovalent ions, such as Cs, Rb, Tl, were also successfully intercalated into the FeSe-layers and then a series of superconductors were discovered [7-10]. Powder X-ray diffraction experiments confirmed that all $AFe_xSe_2$ (A=K, Cs, Rb, Tl) compounds averagely adopt the $ThCr_2Si_2$-type structure and tolerate a striking amount of vacancies in the A- and the Fe-sites. In addition, they share the common features, such as high $T_c$ as high as those of the iron pnictides, ordered Fe vacanices and complex superstructures [11]. More surprisingly, the recent ARPES [12,13] and theoretical calculation indicate $AFe_2Se_2$ only has a clean electron-like Fermi surface, implying the inter-pocket scattering ($S^{\pm}$ pair-symmetry) is absent in the superconducting state [14-16]. The optical spectrum study of $(Tl,K)Fe_{2-x}Se_2$ [17] confirm that the superconductivity originates from the edge of antiferromagnetic (AFM) semiconducting state. It is suggested that the superconductivity is close to AFM Mott-semiconductor from long-range Fe or Fe-vacancies ordered, echoing the behavior of superconductivity in cuprates [18]. In this study, we synthesized an analogous $KFe_xS_2$ layered compound and characterized its crystal structure, electrical and magnetic properties in details. The $KFe_xS_2$ compound exhibits AFM semiconductor ground state, and the superconductivity can be successfully induced by Se-doping. This result supplies new evidence that the superconductivity in ternary iron-chalcogenides is on the verge of Mott localization.

A series of polycrystalline samples were synthesized using a two-step solid state reaction method. First, FeS and FeSSe powders were prepared with iron (Alfa, 99.9+%), sublimed sulfur (98%) and selenium (Alfa, 99.99%) powders, by a similar method described in Ref. 6. Then, the synthesized powders and K (Sinopharm Chemical, 97%) mixed with appropriate stoichiometry were heated in alumina crucibles, sealed in quartz tubes partially backfilled with ultra-high-purity argon. The precursors were put into alumina crucibles and sealed in silica ampoules under an atmosphere of 0.2 bar argon. The samples were heated to 1300 K, cooled down to 1100 K at a rate of 4 K/hr and finally furnace-cooled to room temperature. Well-formed single crystals with shiny surfaces having dimensions up to 5 mm × 5 mm × 0.5 mm could be obtained. The as-prepared samples were characterized by powder X-ray diffraction (PXRD) using a panalytical X'pert diffractometer with Cu $K_a$ radiation. Rietveld refinements of the data were performed with the FULLPROF package [19]. The dc magnetic properties were characterized using a vibrating sample magnetometer (VSM, Quantum Design). The electrical resistivity was measured through the standard four-wire method on the physical property measurement system (PPMS, Quantum Design) with magnetic field up to 9 T. The electric contacts were made using silver paste with the contacting resistance below 0.06 Ω at room temperature. It should be noticed that all physical properties were measured using the obtained crystals.

**Results and discussion**

Figure 1 shows the PXRD pattern and the Rietveld refinement for the as-prepared $KFe_2S_2$ powder sample. Except for a few tiny peaks and iron impurity, all of the XRD reflections can be indexed with a tetragonal unit cell. The experimental results are very similar to those of $KFe_2Se_2$, where the intercalated K ions expel iron out of lattice so as to keep the charge balance. The systemically extinction indicates an average body-centered lattice and the probable space group is *I4/mmm*, which is also similar to $KFe_xSe_2$ and $BaFe_2As_2$. We adopted the structural model of $KFe_2Se_2$ to carry out the Rietveld refinement. The agreement factors were $R_p$=2.03%, $R_{wp}$=2.81%, and $\chi^2$=2.08%, respectively. The refined lattice parameters ($a=b$=3.7451(2) Å, $c$=13.5782(5) Å, and $V$=190.443(2) Å$^3$) are reasonable smaller than those of $KFe_2Se_2$, a consequence of the smaller size of $S^{2-}$ comparing to $Se^{2-}$. The values of the fractional atomic coordinates as well as important bond lengths and bond angles are listed in table Ⅰ. From the crystallographic point of view, the average crystal structure consists of antifluorite-type S-$Fe_X$-S layers forming edge-sharing $Fe_xS_4$ tetrahedra separated by single K cation sheets. Furthermore, there should be ordered iron vacancies in $Fe_xS$ layers, which possibly results in complex superstructure in *ab*- or along *c*-axis. Of particular interest to note is its intraplane Fe-Fe distance is 2.6284(1) Å, which is shorter than 2.7673(5) Å counterparts of $KFe_xSe_2$ [6]. Therefore, the superexchange between the nearest and the next nearest neighbor iron/iron-vacancies may stronger compared with $KFe_xSe_2$, which possibly leads to the antiferromagnetic ground state in $KFe_xS_2$.

Fig. 2 shows the X-ray diffraction patterns of crashed $K_{0.8}Fe_{1.7}SSe$ crystals. No more sulfur and iron impurities were detected and sulfur atoms were all incorporated into the lattice. The lattice parameters expand to $a$=3.8125(3) Å, $c$= 13.9487(3) Å, and

$V$=202.746(2) Å$^3$, which is a natural result compared with above and previous reports [6]. The broader peaks in the powder XRD pattern reflect that there may be layer stacking faults. Like $K_{0.8}Fe_{1.7}Se_2$, it is difficult to observe satellites peaks due to iron vacancies-induced superstructure. Both X-ray diffraction patterns of $K_{0.8}Fe_{1.7}S_2$ and $K_{0.8}Fe_{1.7}SSe$ crystals were plotted in inset of Fig. 2. The patterns only show sharp 00$l$ ($l$=2n) reflections, implying that two samples are perfectly oriented along $c$-axis.

The $ab$-plane electrical resistivity as a function of temperature for the $K_{0.8}Fe_{1.7}S_2$ crystal is displayed in Fig. 3(a). The room temperature electrical resistivity measured is 0.12 Ω·cm, which is comparable to those of $AFe_2Se_2$ [7-10]. As temperature decreases, the resistivity basically obeys thermally activated behavior $\rho(T)=\rho_0\exp(E_a/K_BT)$, where $\rho_0$ refers to a prefactor and $K_B$ is Boltzman's constant. The fitted activation energy gap is about 40 meV using the $\rho(T)$ data from 300 K to 10 K, which is slightly larger than that of $(Tl,K)Fe_{1.5}Se_2$ [10]. The temperature dependence of the susceptibility ($\chi$), measured under a magnetic field of 1 T applied parallel to the $c$-axis, is shown in Fig. 3(b). The susceptibility shows a peak at $T_N$=260 K, implying the occurrence of an antiferromagnetic correlation. Below 260 K, the susceptibility linearly decreases with decreasing temperature. The similar anomaly has been widely observed in $TlFe_xS_2$ [20] and $(Tl,K)Fe_xSe_2$ [10], which may be attributed to the Fe-vacancy ordering.

In order to explore the anion-substitution effect on $K_{0.8}Fe_{1.7}S_2$, we investigated the electrical and magnetic properties of Se-doped $K_{0.8}Fe_{1.7}S_2$ single crystal. Fig. 4(a) represents the $ab$-plane electrical resistivity $\rho(T)$ for the $K_{0.8}Fe_{1.7}SSe$ single crystal. At room temperature, the resistivity of $K_{0.8}Fe_{1.7}SSe$ is 1.13 Ω·cm and this value is about three orders of magnitude larger than those of $FeSe_{1-x}S_x$ [3]. The high resistivity is possible attributed to stronger lattice scattering from the Fe vacancies. The curve qualitatively exhibits a metallic-like behavior in the whole temperature range, being quite different from those of $K_xFe_{1.7}Se_2$ and $Rb_xFe_{1.7}Se_2$, but similar to that of $Cs_xFe_{1.7}Se_2$. Above the superconducting critical temperature, the normal state resistivity monotonously decreases and exhibits a nonlinear trend, which may be correlated to the strong spin scattering effect. It is interesting that there is no hump at the whole measured temperature range. High pressure *in-situ* measurement of $K_{0.8}Fe_{1.7}Se_2$ indicates that the hump coexists with the superconductivity, and it could be suppressed by enhancing the magnitude of ordering [21,22]. Therefore, emergent data are needed to clarify the origin of the hump. With further decreasing temperatures, the superconductivity emerges at about 26.2 K and the resistivity quickly decreases to zero at 24.8 K. By 90/10 criterion, we find the midpoint of the resistive transition where the resistance drops to 50% of that of the onset at 25.6 K. and a transition width of ~1.4 K, suggesting the homogeneous nature of the crystal. The magnetization of $K_{0.8}Fe_{1.7}SSe$ crystal as a function of temperature was shown in Fig. 4(b). Like other FeSe-based superconductors, no more magnetic transitions occur above $T_c$. The ZFC and FC susceptibilities show that the superconducting shield emerges at about 24.5 K with magnetic field of 10 Oe perpendicular to the *ab*-plane, which is consistent with the resistivity measurement. The superconducting volume fraction estimated from the ZFC magnetic at 10 K is about 80%, and the above characterizations indicate the bulk superconductivity nature and good crystallinity of the crystals.

Fig. 5(a) plots the resistivity of $K_{0.8}Fe_{1.7}SSe$ under different magnetic fields applied perpendicular to the *ab*-plane around $T_c$. The superconducting transition temperature is suppressed gradually and the width of transition is broadened with the increase of magnetic filed. In the BCS-theory, the upper critical field at T=0 K can be determined by the Werthamer-Helfand-Hohenberg (WHH) formula $H_{c2}(0)=0.693[-(dH_{c2}/dT)]_{Tc} T_c$ [23]. Using the data of $H_{c2}(T)$ for the 90% resistive transition, a slope $dH_{c2}/dT$=-0.74 T/K can be obtained. The roughly estimated $H_{c2}$ at zero temperature is 13.5 T, which is much smaller compared with those of $AFe_xSe_2$ superconductors [24]. It should be related to the disordered distribution in $Fe_xSe/S$ layers and weak coupling state by the Se-doped.

The intentional Se-doping similarly suppresses the long-range AFM ordering and pushes insulating state into superconducting ground state, which is as same as iron self-doping. It deserves attention that Mott-semiconducting AFM state has been experimentally observed in both $(Tl,K)Fe_{1.5}Se_2$ and $K_xFe_{2+y}Se_2$ [25], and superconductivity can be achieved by varying iron content. Besides, the theoretical calculation also proposed that $Fe_xSe$-base ternary prototype is bi-collinear antiferromagnetic order and Mott insulators [15]. Zhu *et al.* theoretically argued that the expansion of Fe square lattice can result in a narrowing of the 3*d*-electron bands, and finally promotes the correlation effect to Mott localization state [26]. In former works, we speculate that there should have complex superstructure in *ab* plane or along *c*-axis of $Fe_xSe$-based compound, where the larger Fe-square-lattice unit cell would narrow 3*d* bands and favor antiferromagnetically ordered state. However, there still has no apparently physical picture to elucidate how doping destroy the ordered iron/vacancies and further experimental data need to be collected.

In conclusion, we have synthesized a new FeS-based layer compound and studied its physical properties. The electrical and magnetic analyses clearly indicate $K_{0.8}Fe_xS_2$ is an antiferromagnetic semiconductor. We firstly demonstrate that anion Se-doping in $K_{0.8}Fe_{1.7}S_2$ can successfully induce superconductivity at $T_c^{onset} \sim 26.2$ K. It possible gives a clue to better understand the correlation between the superconductivity and Fe/vacancies-ordering in ternary iron-chalcogenides.

This work was partly supported by the National Natural Science Foundation of China under Grants Nos. 90922037, 50872144, and 50972162, the Chinese Academy of Sciences, and the International Centre for Diffraction Data (ICDD, USA).

Table and Figure Captions:

TABLE I. Crystallographic data of KFe$_x$S$_2$

FIG. 1. X-ray diffraction pattern of powdered K$_{0.8}$Fe$_{1.7}$SSe. The inset shows the x-ray diffraction patterns of K$_{0.8}$Fe$_{1.7}$S$_2$ and K$_{0.8}$Fe$_{1.7}$SSe crystal.

FIG. 2 (Color online). (a): In- plane resistivity as a function of temperature for K$_{0.8}$Fe$_{1.7}$S$_2$ crystal. (b): The temperature dependence of magnetic susceptibility measured for K$_{0.8}$Fe$_{1.7}$S$_2$ at 1 Telsa magnetic field with field cooling process.

FIG. 3 (Color online). (a): Temperature dependence of electrical resistivity for K$_{0.8}$Fe$_{1.7}$SSe crystal from 10 to 30 K. (b) The magnetization of K$_{0.8}$Fe$_{1.7}$SSe crystal as a function of temperature with the H parallel to *c*- axis.

FIG. 4 (Color online). (a) The temperature dependence of resistivity for K$_{0.8}$Fe$_{1.7}$SSe crystal with the magnetic field parallel to the c- axis up to 9 T. (b) Temperature dependence of upper critical fields for K$_{0.8}$Fe$_{1.7}$SSe.

| Formula | KFe$_x$S$_2$ |
|---|---|
| Space group | *I4/mmm* |
| $a$ (Å) | 3.7451(2) |
| $c$ (Å) | 13.5782(5) |
| $V$ (Å$^3$) | 190.443(5) |
| Z | 2 |
| $R_p$ | 2.03% |
| $R_{wp}$ | 2.81% |
| $\chi^2$ | 2.08 |
| Atomic parameters: | |
| K | 2$a$ (0, 0, 0) |
| Fe | 4$d$ (0, 0.5, 0.25) |
| Se | 4$e$ (0, 0, $z$) |
| | $z$=0.3504(2) |
| Bond length (Å): | |
| Fe-Se | 2.3462(2) × 4 |
| Fe-Fe | 2.6482(1) × 4 |
| Bond angles (deg): | |
| | 110.267(8) × 4 |
| | 107.89(8) × 2 |

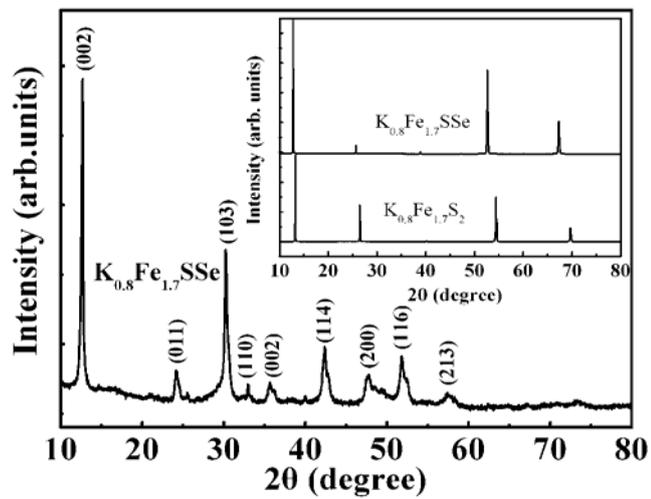

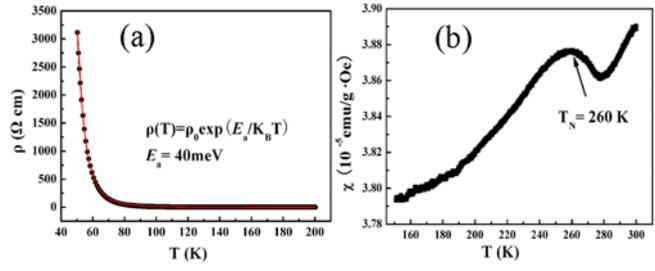

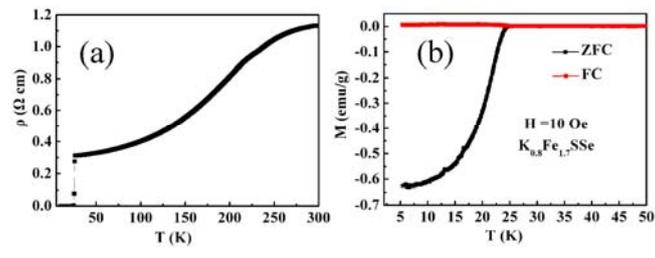

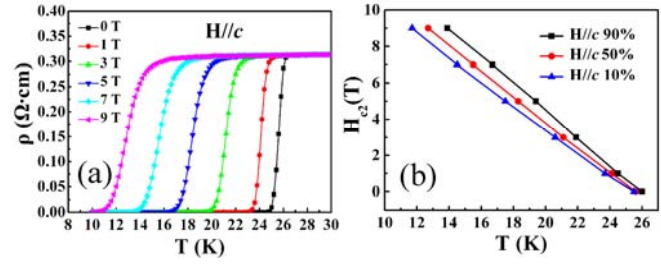